\documentstyle[aps,prl,twocolumn,epsf]{revtex}

\begin{document}

{\bf Comment on ``Tricritical Behavior in Rupture Induced by Disorder" }

In their letter, Andersen, Sornette, and Leung describe possible
behaviors for rupture in disordered media. Their analysis is first based on the
mean field-like democratic fiber bundle model (DFBM). In the
DFBM, $N_0$ ($\rightarrow \infty $) fibers are initially pulled with a force 
$F$, which is distributed uniformly. A fiber breaks if the stress it
undergoes exceeds a threshold chosen from a probability distribution $%
p\left( x\right) =\frac{dP}{dx}$. After each set of failures, the
force $F$ is redistributed over the remaining fibers. Let $N\left( F\right) $
be the final number of intact fibers. What is the behavior of $N\left( F\right) $, as $F$ increases
from $0$ to $\infty $? Ref. \cite{PRL} claims the existence of a tricritical
point, separating a ``first-order'' regime, characterized by a sudden global
failure, from a ``second-order'' regime, characterized by a divergence in
the breaking rate (response function) $N^{\prime }\left( F\right) $.

Here, we present a graphical solution of the DFBM. Unlike an analytical
solution, this enables us to consider the qualitatively different classes of
disorder distribution, and to distinguish the corresponding generic
behaviors of $N\left( F\right) $. We find that, for continuous distributions
with finite mean, the system always undergoes a macroscopic failure,
preceded by a diverging breaking rate. A ``first-order'' failure, with no
preceding divergence, is an artifact of a (large enough) discontinuity 
in $p\left(x\right) $.

Suppose that a set of failures leaves the system with $N_i$ unbroken fibers.
Each of these is now under a stress $F/N_i$. This leads to another set of
failures, bringing the number of intact fibers to $N_{i+1}=N_0\left\{
1-P\left( \frac F{N_i}\right) \right\} $. The function $N\left( F\right) $
defined above is nothing but $N_\infty $. A graphical scheme for this
iteration is facilitated by setting $x_i=N_i/F$, $f=F/N_0$, and $\pi \left(
x\right) =1-P\left( 1/x\right) $, leading to 
\begin{equation}
\label{iteration}fx_{i+1}=\pi \left( x_i\right) \text{.}
\end{equation}
Since
\begin{equation}
\label{slope}\pi ^{\prime }\left( x\right) =\frac 1{x^2}p\left( \frac
1x\right) \text{,}
\end{equation}
$\pi \left( x\right) $ is a monotonic function of $x$, increasing from $0$
to $1$. Therefore, from iterating Eq. (\ref{iteration}) graphically, $%
N\left( F\right) $ is given by the rightmost intersection of the curve $%
y=\pi \left( x\right) $ with the straight line $y=fx$. As the force is
increased, the straight line becomes steeper, and the intersection
consequently moves to the left.

We first consider continuous infinite-support distributions $p\left(
x\right) $. We can distinguish three qualitatively different cases,
depending on the behavior of $p\left( x\right) $ at large $x$ (see Fig. 1). {\it (i) For }%
$p\left( x\right) \sim x^{-r}$, {\it with }$r<2$, intact fibers remain at
any force $F$. {\it (ii) For }$p\left( x\right) =\alpha x^{-2}$, $N\left(
F\right) $ goes continuously to zero at $F_c=N_0\alpha $, with a diverging
breaking rate ($N^{\prime }\left( F_c\right) =\infty $). In both cases,
there may or may not be jumps in $N\left( F\right) $ at smaller forces. In
particular, if the slope of $\pi \left( x\right) $ is monotonically
decreasing ($\pi $ convex), $N\left( F\right) $ has no discontinuity. This
is the case for {\it e.g.} $p\left( x\right) =\frac \alpha {x^r}e^{-\alpha
/x^{r-1}}$ ($1<r\leq 2$), with any $\alpha $. Note also that both classes of
distributions yield an infinite mean $\left\langle x\right\rangle =\infty $. 
{\it (iii) For $p\left( x\right) $ such that }$x^2p\left( x\right)
\rightarrow 0$, {\it e.g.} $p\left( x\right) =\frac x{\lambda
^2}e^{-x/\lambda }$ or $p\left( x\right) =\frac x\lambda e^{-x^2/2\lambda }$ 
{\it with any $\lambda $}, there is at least one jump in $N\left( F\right) $%
, leading to $N=0$. As seen graphically, any such jump is preceded by a
diverging breaking rate, {\it i.e.} the curve $N\left( F\right) $
reaches its discontinuity vertically, generically according to $N^{\prime
}\left( F\right) \sim \left| F-F_c\right| ^{-\frac 12}$.

\begin{figure} 
\epsfxsize=10.5truecm 
\vspace*{-1.6truecm} 
\centerline{
\epsfbox{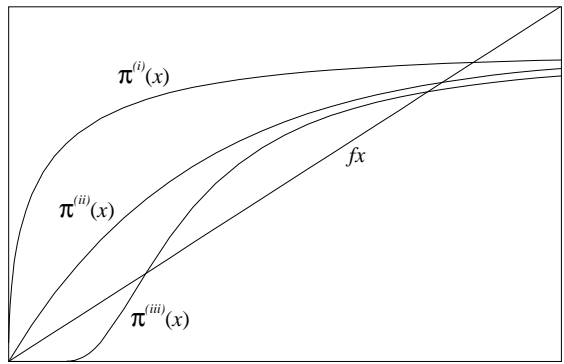} 
} 
\vspace*{-0.7truecm}

\caption{Illustration of the graphical scheme for the three generic cases 
{\it (i)}, {\it (ii)} and {\it (iii)} discussed in the text.}
\label{fig} 
\vspace*{0.1truecm}
\end{figure}

From our graphical method, it clearly appears that a sudden jump in $N\left(
F\right) $ with no divergence preceding it, is possible only if $\pi \left(
x\right) $ has a non-differentiable point, which in turn, by Eq.(\ref{slope}%
), requires a discontinuity in $p\left( x\right) $. We illustrate this in
the context of finite-support distributions, for which $p\left( x\right) =0$
for $x<a$ and $x>b$. No fiber breaks up to $F_a=N_0a$. Then, as can be shown 
by the graphical method, a sudden failure occurs at $F_a$ only if $\pi ^{\prime }\left(
1/a\right) \geq a$, which is equivalent to requiring a minimal discontinuity 
$p\left( a\right) \geq 1/a$ at $a$. Also, note that for any finite $b$,
there will be at least one jump in $N\left( F\right) $, leading to $N=0$.
Thus, for the more physical case of a continuous - albeit finite-support -
distribution, the behavior of the solution is identical to that of case (%
{\it iii}) above.

In summary, the generic form of the DFBM's solution depends on the disorder
distribution {\it only via }its large $x$ behavior - and possible
discontinuity points.

I would like to thank Professor Mehran Kardar for fruitful discussions. This
work was partly supported by the NSF through grant No. DMR-93-03667.

Rav\'a da Silveira

Department of Physics

Massachusetts Institute of Technology

Cambridge, Massachusetts 02139

\end{document}